\documentclass[twocolumn,aps,nofootinbib]{revtex4}

\usepackage{graphicx}
\usepackage{epstopdf}
\usepackage{latexsym}
\usepackage{amssymb}
\usepackage{color}

\usepackage[center]{subfigure}

\begin{document}

 \newcommand{\bq}{\begin{equation}}
 \newcommand{\eq}{\end{equation}}
 \newcommand{\bqn}{\begin{eqnarray}}
 \newcommand{\eqn}{\end{eqnarray}}
 \newcommand{\nb}{\nonumber}
 \newcommand{\lb}{\label}
\newcommand{\PRL}{Phys. Rev. Lett.}
\newcommand{\PL}{Phys. Lett.}
\newcommand{\PR}{Phys. Rev.}
\newcommand{\CQG}{Class. Quantum Grav.}
\newcommand{\Alan}[1]{\noindent{\color{red} #1 }}
\newcommand{\duvida}[1]{\noindent{\color{blue} #1 }}
\newcommand{\Kai}[1]{\noindent{\color{green} #1 }}

\title{Scalar Quasinormal Modes of Anti-de Sitter Static Spacetime in Horava-Lifshitz Gravity with $U(1)$ Symmetry}

\author{Kai Lin$^{1,2)}$}\email{lk314159@hotmail.com}
\author{Wei-Liang Qian$^{3,4)}$}\email{wlqian@usp.br}
\author{A. B. Pavan$^{1)}$}\email{alan@unifei.edu.br}

\affiliation {1) Universidade Federal de Itajub\'a, Instituto de F\'isica e Qu\'imica, Itajub\'a, Brazil}
\affiliation{2) Instituto de F\'isica, Universidade de S\~ao Paulo, S\~ao Paulo, Brazil}
\affiliation{3) Escola de Engenharia de Lorena, Universidade de S\~ao Paulo, Lorena, SP, Brasil}
\affiliation{4) Faculadade de Engenharia de Guaratinguet\'a, Universidade Estadual Paulista, Guaratinguet\'a, SP, Brasil}

\date{\today}

\begin{abstract}
In this paper, we investigate the scalar quasinormal modes of
Ho\v{r}ava-Lifshitz theory with $U(1)$ symmetry in static Anti-de
Sitter spacetime. The static planar and spherical black hole
solutions in lower energy limit are derived in non-projectable
Ho\v{r}ava-Lifshitz gravity. The equation of motion of a scalar
field is obtained, and is utilized to study the quasinormal modes of
massless scalar particles. We find that the effect of
Ho\v{r}ava-Lifshitz correction is to increase the quasinormal period
as well as to slow down the decay of the oscillation magnitude.
Besides, the scalar field could be unstable when the correction
becomes too large.
\end{abstract}


\maketitle

\section{Introduction}
\renewcommand{\theequation}{1.\arabic{equation}} \setcounter{equation}{0}

Einstein's general relativity provides a unified description of gravity as a geometric property of space and time.
It is the simplest theory that is consistent with the experimental data at the highest achievable precision to date.
However, several unanswered questions remain, and the most fundamental one is how general relativity could accommodate to the quantum field theory to produce a complete theory of quantum gravity.
Ho\v{r}ava-Lifshitz theory \cite{HL} (HL) was proposed by Ho\v{r}ava in 2009 as a renormalizable quantum gravity candidate.
General relativity is restored as the infrared limit of the theory, and any deviation would remain suppressed under current experimental constraints.
However, the Lorentz symmetry is broken in the ultraviolet sector,  the anisotropic scaling of space and time being given by
 \bqn
 \label{INT1}
\textbf{x}\rightarrow \ell\textbf{x},~~~~~~~t\rightarrow \ell^z t,
 \eqn
and the $4$-dimensional diffeomorphism is replaced by the foliation-preserving diffeomorphism Diff$({\cal M, F})$,  obeying the transformation
 \bqn
 \label{INT2}
\textbf{x}\rightarrow \textbf{x}'(t,\textbf{x}),~~~~~~~t\rightarrow  t'(t).
 \eqn
In 3+1 dimensional spacetime, a power-counting renormalizable
gravity theory must satisfy $z\geq 3$, and the dispersion relation
generically takes the form
 \bqn
 \label{INT3}
E^2=c_p^2p^2\left(1+\sum_{n}\alpha_n\left(\frac{p}{M_*}\right)^{2n}\right),
 \eqn
where $p$ and $E$ are the momentum and energy of the particle, and
$c_p$ is the speed of light at the infrared (IR) limit. $M_*$ denotes
the suppression energy scale of the higher order operators, and the
higher order terms with coefficient $\alpha_n$ are dominant at the
ultraviolet (UV) case. The aim of Ho\v{r}ava-Lifshitz theory is to
construct a renormalizable gravity, which for the $z=3$ case
satisfies the condition of renormalization, therefore it is not necessary
to consider the higher order terms with $n=1,2$ in
Ho\v{r}ava-Lifshitz gravity.

Ho\v{r}ava-Lifshitz gravity has attracted the attention of many physicists.
In particular, topics such as ghost modes, instability, strong coupling and exceeding degrees of freedom have intrigued many studies.
In order to tackle these difficulties, Ho\v{r}ava et al proposed projectability, detailed balance, and in particular, an extra local U(1) symmetry  satisfying $U(1) \ltimes {\mbox{Diff}}(M, \; {\cal{F}})$ \cite{HM}.
Recent developments following this line of thought have made further improvements \cite{Work,WorkI,AM,PostNewtonian}.

Another important question that can bring additional information about this theory is the stability of black hole solutions. Stability is a central question to be dealt with
considering gravitational solutions, which might be either physically irrelevant or might lead to phase transitions when unstable \cite{zpwa,apop}.  It can be addressed studying their quasinormal modes since they describe exponentially decreasing oscillation in time while the black hole evolves towards the perfect spherical shape. These modes provide valuable information on the main properties of black hole. In addition, the quasinormal modes of matter fields evolving near to the event horizon can also provide information about the spacetime. Moreover, it is interesting to study the quasinormal modes of Anti-de Sitter black hole, because of its important implications in the context of the AdS/CFT correspondence.

This work involves an attempt to investigate the scalar quasinormal
modes of Ho\v{r}ava-Lifshitz gravity with $U(1)$ symmetry in static
Anti-de Sitter spacetime at infrared (IR) limit by ignoring higher order correction terms in the theory (such as ${\cal L}^H_V$ in Eq.(2.4) and ${\cal L}_M^H$ in Eq.(3.4), see the text below).
As a result, the dispersion relation at IR limit remains the same form as that in
general relativity:
 \bqn
 \label{INT4}
E^2=c_p^2p^2.
 \eqn
In this context, we study the quasinormal modes of massless scalar
particles with velocity of light. It implies that the boundary
condition at the killing event horizon is a pure in-going mode, and
therefore, for the massless scalar field investigated in this paper,
the Ho\v{r}ava-Lifshitz Anti-de Sitter black hole is defined by
event horizon at IR limit.

The outline of the present paper is as follows.
In section II, we study the non-projectable Ho\v{r}ava-Lifshitz Gravity with $U(1)$ symmetry, and obtain the static planar and spherical black hole solutions in the lower energy limit.
The equation for scalar quasinormal modes is subsequently derived in section III.
The dynamical properties of the quasinormal modes are investigated in section IV and V.
Section VI is dedicated to conclusion remarks.
We relegate the results of the static planar black hole solutions in the projectable Ho\v{r}ava-Lifshitz Gravity with local $U(1)$ symmetry to Appendix.

\section{non-Projectable Solutions in HL theory}
\renewcommand{\theequation}{2.\arabic{equation}} \setcounter{equation}{0}

Here we derive the planar and spherical black hole solutions in non-projectable Ho\v{r}ava-Lifshitz theory with local $U(1)$ symmetry.

The action reads \cite{WorkI,ZWWS,LMWZ},
\bqn
 \label{action1}
S&=&\zeta^2 \int dtd^3x \sqrt{g} N \Big({\cal{L}}_{K}-{\cal{L}}_{V} + {\cal{L}}_{A}
+ {\cal{L}}_{\varphi}\nb\\
&& ~~~~~~~~~~~~~~~~~~~~~~~+{\cal{L}}_S  + {\zeta^{-2}} {\cal{L}}_M\Big),
 \eqn
where $g={\rm det}(g_{ij}),\; \zeta^2 \equiv
1/(16\pi G)$ with $G$ being the Newtonian constant of the theory. Matter fields are introduced through ${\cal{L}}_M$ and
 \bqn
 \lb{L0}
{\cal{L}}_{K} &=& K_{ij}K^{ij} -   \lambda K^{2},\nb\\
{\cal{L}}_{A} &=&\frac{A}{N}\Big(2\Lambda_{g} - R\Big), \nb\\
{\cal{L}}_{\varphi} &=&  \varphi{\cal{G}}^{ij}\big(2K_{ij}+\nabla_i\nabla_j\varphi+a_i\nabla_j\varphi\big)\nb\\
& & +(1-\lambda)\Big[\big(\Delta\varphi+a_i\nabla^i\varphi\big)^2
+2\big(\Delta\varphi+a_i\nabla^i\varphi\big)K\Big]\nb\\
& & +\frac{1}{3}\hat{\cal G}^{ijlk}\Big[4\left(\nabla_{i}\nabla_{j}\varphi\right) a_{(k}\nabla_{l)}\varphi \nb\\
&&  ~~ + 5 \left(a_{(i}\nabla_{j)}\varphi\right) a_{(k}\nabla_{l)}\varphi + 2 \left(\nabla_{(i}\varphi\right)a_{j)(k}\nabla_{l)}\varphi \nb\\
&&~~ + 6K_{ij} a_{(l}\nabla_{k)}\varphi \Big],\nb\\
{\cal{L}}_S &=&\frac{A - {\cal{A}}}{N}  (\sigma_1 a^ia_i+\sigma_2a^i_{\;\;i}),
 \eqn
where $N$, $N_i$ and $g_{ij}$ are, the lapse function, shift vector, and 3-metric with at fixed $t$ in the ADM decomposition, respectively and $i$ runs from 1 to 3 in spatial coordinates. The functions $A$ and $\varphi$ are the gauge field and Newtonian prepotential.

Besides, $\Delta \equiv g^{ij}\nabla_{i}\nabla_{j}$ and  $ \hat{\cal G}^{ijlk} =  g^{il}g^{jk} - g^{ij}g^{kl}$,  while $\Lambda_{g}$ is a coupling constant. The Ricci scalar  and tensor are $R = g^{ij}R_{ij}$ and $R_{ij} = g^{kl}R_{kilj}$. The Riemann tensor $R_{i jkl}$ is
 \bqn
 \lb{L1}
 R_{ijkl} &=& g_{ik}R_{jl}   +  g_{jl}R_{ik}  -  g_{jk}R_{il}  -  g_{il}R_{jk}\nb\\
 &&    - \frac{1}{2}\left(g_{ik}g_{jl} - g_{il}g_{jk}\right)R,\nb\\
K_{ij} &\equiv& \frac{1}{2N}\left(- \partial_t{g}_{ij} + \nabla_{i}N_{j} +
\nabla_{j}N_{i}\right),\nb\\
{\cal{G}}_{ij} &\equiv& R_{ij} - \frac{1}{2}g_{ij}R + \Lambda_{g} g_{ij},\nb\\
a_{i} &\equiv& \frac{N_{,i}}{N},\;\;\; a_{ij} \equiv \nabla_{j} a_{i},\nb\\
{\cal A}&\equiv &-\partial_t\varphi+N^i\nabla_i\varphi+\frac{N}{2}\left(\nabla_i\varphi\right)\left(\nabla^i\varphi\right).
 \eqn
For simplicity, in this work we only consider the case where $\sigma_1=\sigma_2=0$ and $\lambda=1$.

All higher order corrections in Ho\v{r}ava-Lifshitz theory ${\cal
L}_V^H$ are included in ${\cal L}_V$ shown in
\cite{WorkI,PostNewtonian}, which could be written as
 \bqn
 \lb{L2}
{\cal L}_V=2\Lambda-\beta_0a_ia^i+\gamma_1 R+{\cal L}^H_V,
 \eqn
Again for simplicity, we take $\Lambda=\Lambda_g$ and ${\cal{L}_M} =0$. For the lower energy case, we also ignore the contribution from ${\cal L}^H_V$ since it is a higher order correction. Thus, implementing the restrictions above mentioned the resultant action to be addressed is
 \bqn
 \label{action2}
S&=&\zeta^2 \int dtd^3x \sqrt{g} N \Big({\cal{L}}_{K}-{\cal{L}}_{V} + {\cal{L}}_{A}
+ {\cal{L}}_{\varphi}\Big)
 \eqn

By making use of the action (\ref{action2}), one proceeds to derive the field equations for planar and spherical black hole spacetime \cite{LMWZ}.

First, let us consider the planar black hole spacetime, whose metric is
 \bqn
 \lb{BS1}
ds^2=-f(r)dt^2+\frac{\left[dr+h(r)dt\right]^2}{f(r)}+r^2(dx^2+dy^2),
 \eqn
where $f(r)=N^2$ and $h(r)=N^i$. By substituting the metric into
field equations, one obtains the following three independent field
equations
 \bqn
 \lb{BS2}
rf'+f+\Lambda_gr^2=0,
 \eqn
 \bqn
 \lb{BS3}
&&\beta_0\left(\frac{f''}{f}-\frac{f'^2}{4f^2}+\frac{2f'}{rf}\right)-\frac{4hh'}{rf^2}+\left(1+\frac{h^2}{f^2}\right)\frac{2f'}{rf}\nb\\
&&~~~~~~-\frac{2h^2}{r^2f^2}+\frac{2}{r^2}+\frac{2\Lambda_g}{f}=0,
 \eqn
 \bqn
 \lb{BS4}
&&A'+\left(\frac{1}{r}+\frac{r\Lambda_g}{f}\right)\frac{A}{2}-\frac{\sqrt{f}}{2r}-\left(1+\frac{h^2}{f^2}\right)\frac{f'}{2\sqrt{f}}\nb\\
&&-\frac{r\Lambda_g}{2\sqrt{f}}-\frac{\beta_0rf'^2}{16f^{3/2}}+\frac{h^2}{2rf^{3/2}}+\frac{hh'}{f^{3/2}}=0 \,,
 \eqn
whose solutions read
 \bqn
 \lb{BS5}
f&=&f_{PBH}=-\frac{\Lambda_g}{3}r^2-\frac{M}{r},\nb\\
h^2&=&h_0\frac{f}{r}+\frac{\beta_0f}{8}\left[\frac{M}{r}\ln\left|\frac{1}{27r^2f^3}\right|-\frac{20}{9}\Lambda_gr^2\right],\nb\\
A&=&\frac{\beta_0}{8}\left[3\sqrt{f}+\frac{\Lambda_gr^2}{\sqrt{f}}+\sqrt{f}\ln\left|\frac{1}{27r^2f^3}\right|\right]\nb\\
&&+A_0\sqrt{f},
 \eqn
where $M$, $A_0$ and $h_0$ are constants.

Next, let us consider the metric for spherical black hole
 \bqn
 \lb{BH1}
ds^2&=&-f(r)dt^2+\frac{\left[dr+h(r)dt\right]^2}{f(r)}\nb\\
&&+r^2(d\theta^2+\sin^2\theta d\phi^2),
 \eqn
which again leads to three field equations as
 \bqn
 \lb{BH2}
rf'+f+\Lambda_gr^2=1,
 \eqn
 \bqn
 \lb{BH3}
&&\beta_0\left(\frac{f''}{f}-\frac{f'^2}{4f^2}+\frac{2f'}{rf}\right)-\frac{4hh'}{rf^2}+\left(1+\frac{h^2}{f^2}\right)\frac{2f'}{rf}\nb\\
&&~~~~~~-\frac{2h^2}{r^2f^2}+\frac{2}{r^2}\left(1-\frac{1}{f}\right)+\frac{2\Lambda_g}{f}=0,
 \eqn
 \bqn
 \lb{BH4}
&&A'+\left(\frac{1}{r}-\frac{1}{rf}+\frac{r\Lambda_g}{f}\right)\frac{A}{2}-\left(1+\frac{h^2}{f^2}\right)\frac{f'}{2\sqrt{f}}\nb\\
&&-\frac{r\Lambda_g}{2\sqrt{f}}-\frac{f-1}{2r\sqrt{f}}-\frac{\beta_0rf'^2}{16f^{3/2}}+\frac{h^2}{2rf^{3/2}}+\frac{hh'}{f^{3/2}}=0. \nb\\
 \eqn
By integrating Eq.(\ref{BH2}), we get
 \bqn
 \lb{BH5}
f=f_{SBH}=1-\frac{2M}{r}-\frac{\Lambda_g}{3}r^2.
 \eqn
If we consider this expression to be a black hole solution, it is convenient to rewrite the above equation as
 \bqn
 \lb{BH6}
f=-\frac{\Lambda_g}{3}\left(1-\frac{r_0}{r}\right)\left(r^2+r_0r+r_0^2-\frac{3}{\Lambda_g}\right)\quad ,
 \eqn
where $r_0$ is the position of event horizon. In terms of $r_0$, the solutions for $A$ and $h$ can be expressed as
 \bqn
 \lb{BH7}
A&=&A_0\sqrt{f}+\beta_0\sqrt{f}\big[8\sqrt{3}(r-r_0)\sqrt{r_0^2\Lambda_g-4}(r_0^2\Lambda_g\nb\\
&&-1)(r_0^2\Lambda_g+rr_0\Lambda_g+r^2\Lambda_g-3)\big]^{-1}\nb\\
&&\times\Big\{6 r_0 \sqrt{\Lambda _g} (r^3 \Lambda _g-r_0^3 \Lambda _g-3 r+3 r_0)\nb\\
&&\times \arctan\left(\frac{\left(2 r+r_0\right) \sqrt{\Lambda_g}}{\sqrt{3} \sqrt{r_0^2 \Lambda _g-4}}\right)\nb\\
&&+\sqrt{3} \sqrt{r_0^2 \Lambda _g-4} \Big[-r_0^5 \Lambda _g^2 \Big[\ln (r)\nb\\
&&-3 \big(\ln \big(3-\left(r^2+r_0 r+r_0^2\right) \Lambda _g\big)+\ln \left(r-r_0\right)\nb\\
&&-1\big)\Big]+r_0^3 \Lambda _g \big[-13 \log \left(3-\left(r^2+r_0 r+r_0^2\right) \Lambda _g\right)\nb\\
&&+4 \ln(r)-10 \ln \left(r-r_0\right)+12\big]\nb\\
&&+r r_0^2 \Lambda _g \Big[-r^2 \Lambda _g \big[3 \Big(\ln \left(3-\left(r^2+r_0r+r_0^2\right) \Lambda _g\right)\nb\\
&&+\ln\left(r-r_0\right)\Big)-\ln(r)\big]+9 \ln (3-(r^2+r_0 r\nb\\
&&+r_0^2) \Lambda _g)-3 \ln (r)+9 \ln \left(r-r_0\right)-6\Big]\nb\\
&&-3 r_0 \Big[-4 \ln \left(3-\left(r^2+r_0r+r_0^2\right) \Lambda _g\right)+\ln (r)\nb\\
&&-\ln\left(r-r_0\right)+3\Big]+r \big[r^2 \Lambda _g \big(4 \ln(3-(r^2+r_0 r\nb\\
&&+r_0^2) \Lambda _g)-\ln (r)+\ln\left(r-r_0\right)\big)\nb\\
&&+3 \big[-4 \ln \left(3-\left(r^2+r_0r+r_0^2\right) \Lambda _g\right)+\ln (r)\nb\\
&&-\ln\left(r-r_0\right)+2\big]\big]\Big]\Big\},\nb\\
h^2&=&h_0\frac{f}{r}+\frac{\beta_0(r-r_0)}{216\sqrt{\Lambda_g}r^2}(r_0^2\Lambda_g+rr_0\Lambda_g+r^2\Lambda_g\nb\\
&&-3)\Big\{\sqrt{\Lambda _g} \Big[20 r^3 \Lambda _g+36 r_0 \ln \big(r^2\left(-\Lambda _g\right)\nb\\
&&-r_0 r \Lambda _g-r_0^2 \Lambda _g+3\big)-9r_0^3 \Lambda _g \ln \big(r^2 \left(-\Lambda _g\right)\nb\\
&&-r_0 r \Lambda_g-r_0^2 \Lambda _g+3\big)+3 r_0 \ln(r) \left(r_0^2 \Lambda_g-3\right)\nb\\
&&+\ln \left(r-r_0\right) \left(9 r_0-9 r_0^3 \Lambda_g\right)-36 r\Big]\nb\\
&&+18 \sqrt{3} \sqrt{r_0^2 \Lambda _g-4} \arctan\left(\frac{\left(2 r+r_0\right) \sqrt{\Lambda _g}}{\sqrt{3}
   \sqrt{r_0^2 \Lambda _g-4}}\right)\Big\}.\nb\\
 \eqn
We note that the black hole solutions in general relativity can be viewed as a special case of the above solutions.

\section{Perturbation Equation for Scalar Field}
\renewcommand{\theequation}{3.\arabic{equation}} \setcounter{equation}{0}

We intend to study asymptotically Anti-de Sitter planar and spherical black holes.
For simplicity, we also choose $h_0=0$ and $\beta_0 =0$ so that the resulting solutions possess the Schwarzschild form in the low energy limit \cite{LMWZ}:
 \bqn
 \lb{Metric1}
f_{PBH}(r)&=&r^2-\frac{r_0^3}{r},\nb\\
f_{SBH}(r)&=&\left(1-\frac{r_0}{r}\right)\left(r^2+r_0r+r_0^2+1\right).
 \eqn
with the functions
 \bqn
 \lb{Metric2}
N&=&\sqrt{f(r)},~~N_i=0,~~\Lambda_g=-3, \nonumber\\
A&=&A_0\sqrt{f(r)},~~~\varphi=0,~~h_0=\beta_0=0.
 \eqn
Under these conditions the metrics (\ref{BS1}) and (\ref{BH1}) become diagonal and can be written as
\bqn
\lb{Metric3}
ds^2=-f(r)\ dt^2+\frac{dr^2}{f(r)}+r^2d\sigma^2
\eqn
where $d\sigma^2=dx^2+dy^2$ for planar black hole and $d\sigma^2=d\theta^2+\sin^2\theta d\phi^2$ for spherical black hole, respectively.
Obviously, the Ho\v{r}ava-Lifshitz correction comes from the $A(r)$ term, and these black hole solutions become Schwarzschild Anti-de Sitter solutions when $A_0=0$.

In the lower energy limit, the dispersion relation of general
relativity is recovered. We study the quasinormal modes of massless
scalar particles with velocity of light, so the event horizon is
defined by $f(r_0)=0$, where the event horizon of the black hole is
$r_0$, and the temperature of the black hole is
$T=\frac{f'(r_0)}{4\pi}$.

On the other hand, in the Ho\v{r}ava-Lifshitz theory with $U(1)$ symmetry, the action of the scalar field is given by \cite{AM,HLALW}
 \bqn
 \lb{Matter1}
{\cal L}_M&=&\frac{1}{2N^2}\left[\partial_t\Psi-\left(N^i+N\nabla^i\varphi\right)\nabla_i\Psi\right]^2\nb\\
&&-\left(\frac{1}{2}-\alpha_2\right)\nabla_i\Psi\nabla^i\Psi-\frac{m^2}{2}\Psi^2\nb\\
&&+\frac{A-{\cal A}}{N}\left[c_1\Psi\Delta\Psi+c_2\nabla_i\Psi\nabla^i\Psi\right]+{\cal L}^{H}_{M} \label{lmatter}
 \eqn
where $\alpha_2$, $c_1$ and $c_2$ are arbitrary
functions of the scalar field $\Psi$;
and ${\cal L}_M^H$ comes from higher order corrections in Ho\v{r}ava-Lifshitz theory.
Once we treat the scalar field as a perturbation in the background we shall discard these higher order corrections.
We note that $A=A_0\sqrt{f}=A_0N(r)$, and therefore
 \bqn
 \lb{simplify1}
&&\frac{A-{\cal
A}}{N}\left[c_1\Psi\Delta\Psi+c_2\nabla_i\Psi\nabla^i\Psi\right]-\left(\frac{1}{2}-\alpha_2\right)\nabla_i\Psi\nabla^i\Psi\nb\\
&&=-\left(\frac{1}{2}-\alpha_2-A_0c_2\right)\nabla_i\Psi\nabla^i\Psi+A_0c_1\Psi\Delta\Psi,
 \eqn
Thus we can carry out the substitutions $\alpha_2\rightarrow\alpha_2-A_0c_2+c_0$
and $c_1\rightarrow\frac{c_0}{A_0}$ on the right hand side of the above expression, which is equivalent to choose $c_1=c_2 \equiv c_0$ and $A_0=1$ in Eq(\ref{lmatter}).

Substituting in the action the Eqs.(\ref{Metric1}, \ref{Metric2}, \ref{Metric3}) and setting $\Psi=\frac{\Phi(t,r)}{r}Y(x^k)$ (where $x^k$ is the angular part of the spatial coordinates), we finally get the radial scalar field equation as
 \bqn
 \lb{Fequation1}
\frac{\partial^2\Phi}{\partial {r_*^2}}-(1-2\alpha_2)^{-1}\frac{\partial^2\Phi}{\partial t^2}=V(r)\Phi,
 \eqn
where $r_*=\int \frac{dr}{f(r)}$, and
 \bqn
 \lb{Fequation2}
V(r)&=&f \left\{ \mu^2+\frac{\kappa}{r^2}+\frac{f'}{r} \right.\nb\\
&& \left. -\rho\sqrt{f}\left[A''+\frac{2}{r}A'+\frac{f'}{2f}A'\right]\right\},
 \eqn
where $\mu^2=m^2/(1-2\alpha_2)$ is the effective mass of the scalar field, and $c_1=c_2=\rho(1-2\alpha_2)$.
We note that the second to last term involving $\rho$ in Eq.(\ref{lmatter}) appears due to the local U(1) symmetry,
which was introduced to fulfill the physical requirements to achieve the cancelation of strong coupling, ghost free, stability as well as a reasonable number of coupling constants \cite{AM,HLALW}.
Therefore, the constant $\rho$ measures the strength of the coupling between the scalar field $\Psi$ and U(1) gauge field $A$.
As shown below, it turns out that the stability of the quasi normal modes depends crucially on the value of this parameter.
$\kappa$ is a constant determined by angular part of the scalar field equation. In particular, we have $\kappa=2L^2$ for planar black hole and $\kappa=L(L+1)$ for spherical black hole case, where $L=0,1,2,3,...$ is the azimuthal quantum number.

One may rescale the time $t\rightarrow t/\sqrt{1-2\alpha_2}$ in the above equation, and set $\Phi=e^{-i\omega t}\Phi_0(r)$, so that Eq.(\ref{Fequation1}) can be rewritten as
 \bqn
 \lb{Fequation3}
\frac{\partial^2\Phi_0}{\partial {r_*^2}}+(\omega^2-V)\Phi_0=0,
 \eqn
which is the form of the perturbation equation we use to study quasinormal modes of the scalar field.

\section{Quasinomal Modes by Horowitz-Hubeny Method}
\renewcommand{\theequation}{4.\arabic{equation}} \setcounter{equation}{0}

In this section, we use the method proposed by Horowitz and Hubeny \cite{HH} to calculate the quasinormal modes of a massless ($m=0$) scalar field in the black hole solution discussed above. By introducing $\Phi_0=e^{-i\omega r_*}\psi$, Eq.(\ref{Fequation3}) becomes
 \bqn
 \lb{FequationH1}
f\frac{d^2\psi}{dr^2}-(2i\omega-f')\frac{d\psi}{dr}-V_r(r)\psi=0,
 \eqn
where $V_r(r)=V(r)/f(r)$. By using the transformation $x=1/r$, the above equation can be rewritten into
 \bqn
 \lb{FequationH2}
s(x)\frac{d^2\psi}{dx^2}-\frac{t(x)}{x-x_+}\frac{d\psi}{dx}-\frac{u(x)}{(x-x_+)^2}\psi=0,
 \eqn
where $x_+=1/r_0$ and
 \bqn
 \lb{FequationH3}
s(x)&=&-\frac{x^4f}{x-x_+},\nb\\
t(x)&=&-x^2\left(x^2\frac{df}{dx}+2xf+2i\omega\right),\nb\\
u(x)&=&(x-x_+)V_r,
 \eqn
Expanding $s(x)$, $t(x)$, $u(x)$ and $\psi(x)$ as
 \bqn
 \lb{FequationH4}
s(x)&=&\sum\limits^\infty_{n=0}s_n(x-x_+)^n,\nb\\
t(x)&=&\sum\limits^\infty_{n=0}t_n(x-x_+)^n,\nb\\
u(x)&=&\sum\limits^\infty_{n=0}u_n(x-x_+)^n,\nb\\
\psi(x)&=&\sum\limits^\infty_{n=0}a_n(x-x_+)^n,
 \eqn
and substituting (\ref{FequationH4}) into (\ref{FequationH2}), we get a recursive relation
 \bqn
 \lb{FequationH5}
a_n=-\frac{1}{P_n}\sum\limits^{n-1}_{k=0}\left[k(k-1)s_{n-k}+kt_{n-k}+u_{n-k}\right]a_k.
 \eqn
On the other hand, the boundary condition requires $\psi$ to be purely ingoing mode at the horizon, while it vanishes at infinity. So we let $a_0=1$ and the $a_n$ satisfy the relation
  \bqn
 \lb{FequationH6}
\sum\limits^\infty_{n=0}a_n(-x_+)^n=0.
 \eqn
In what follows, we proceed to evaluate $\omega$ from the above equation.

\begin{figure*}
\includegraphics[width=5cm]{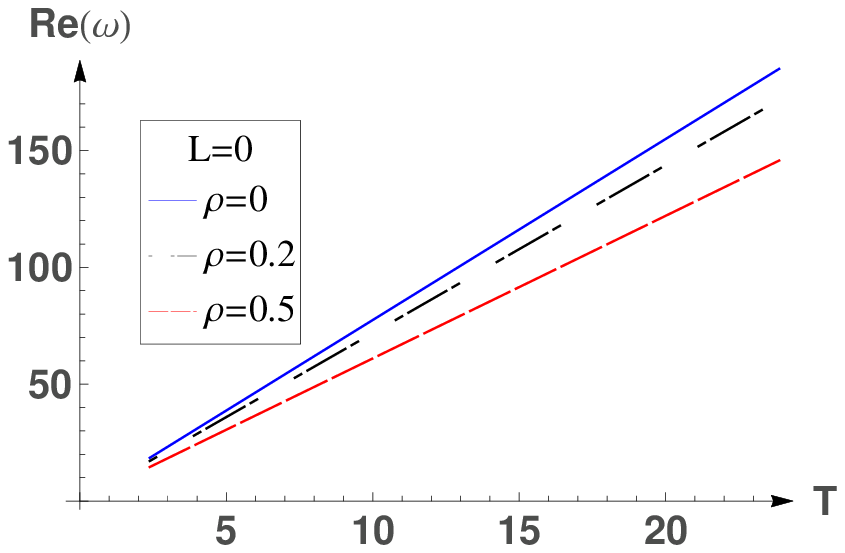}\includegraphics[width=5cm]{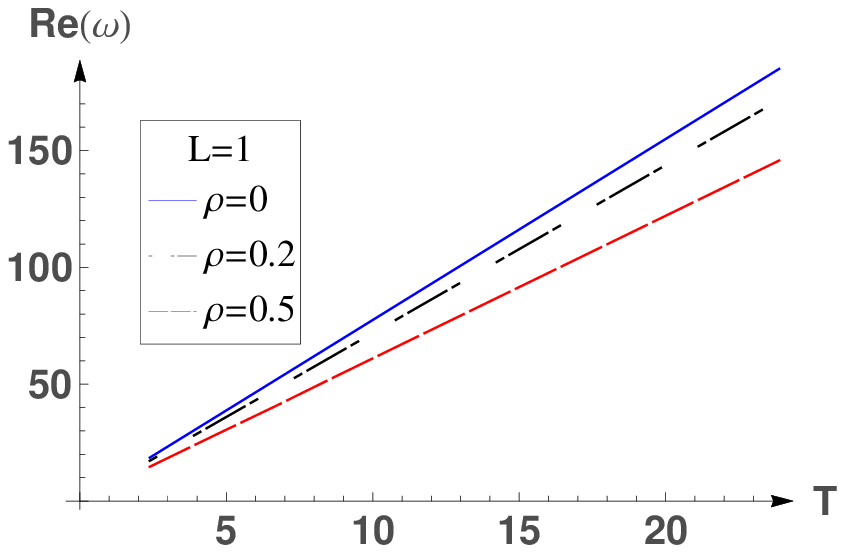}\includegraphics[width=5cm]{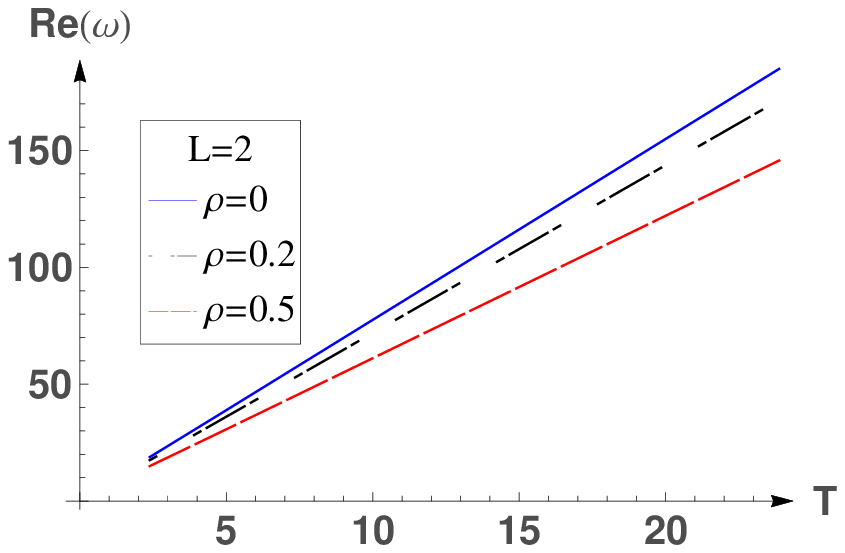}\protect\\
\includegraphics[width=5cm]{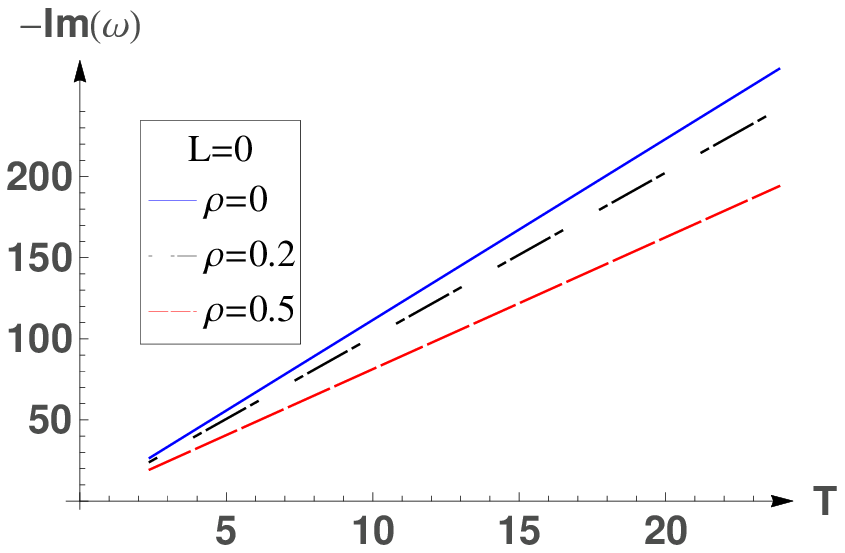}\includegraphics[width=5cm]{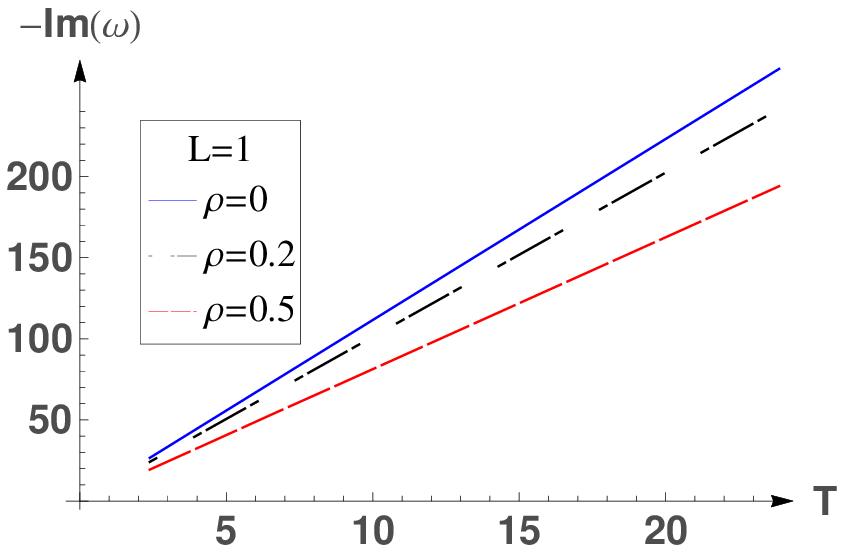}\includegraphics[width=5cm]{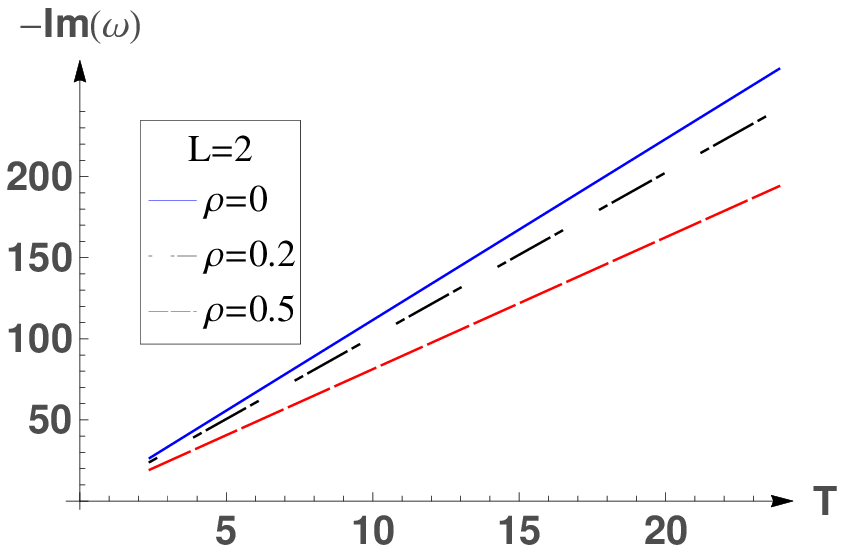}
\caption{Scalar quasinormal modes vs. temperature of
Ho\v{r}ava-Lifshitz planar black hole spacetime, where
$T=\frac{3r_0}{4\pi}$} \label{figA}
\end{figure*}

\begin{figure*}
\includegraphics[width=5cm]{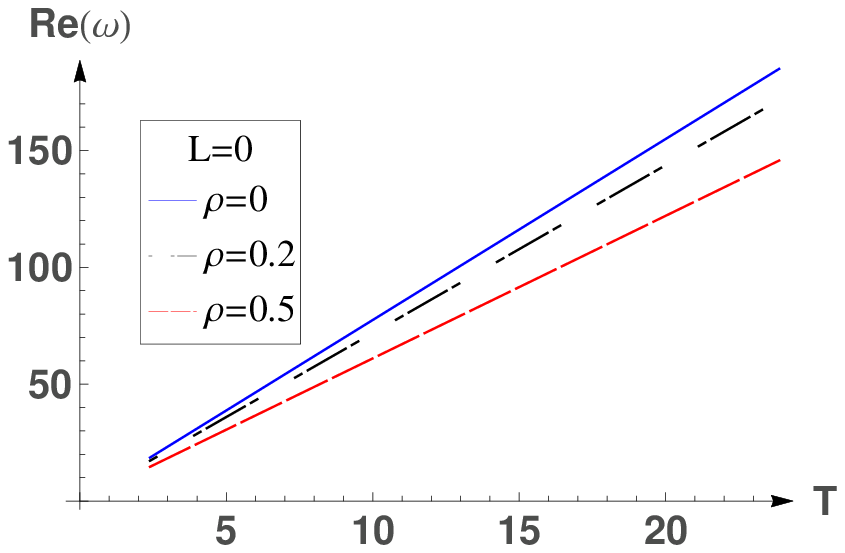}\includegraphics[width=5cm]{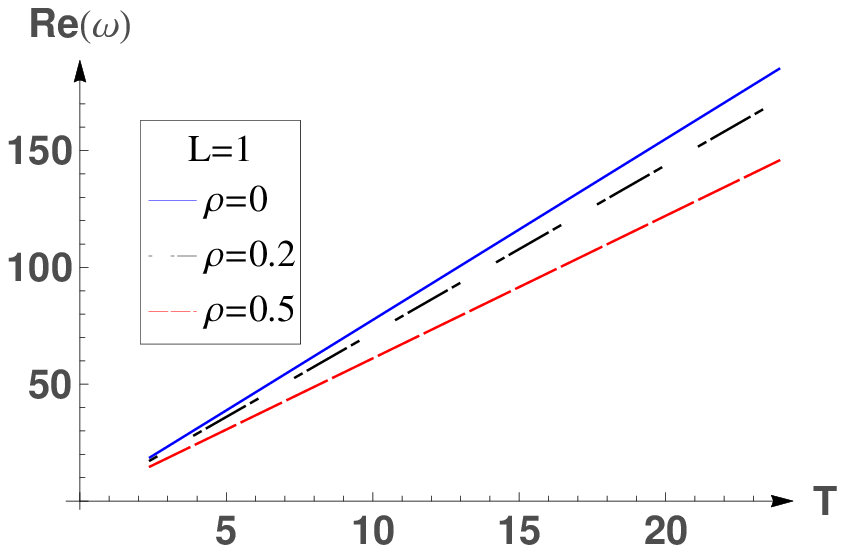}\includegraphics[width=5cm]{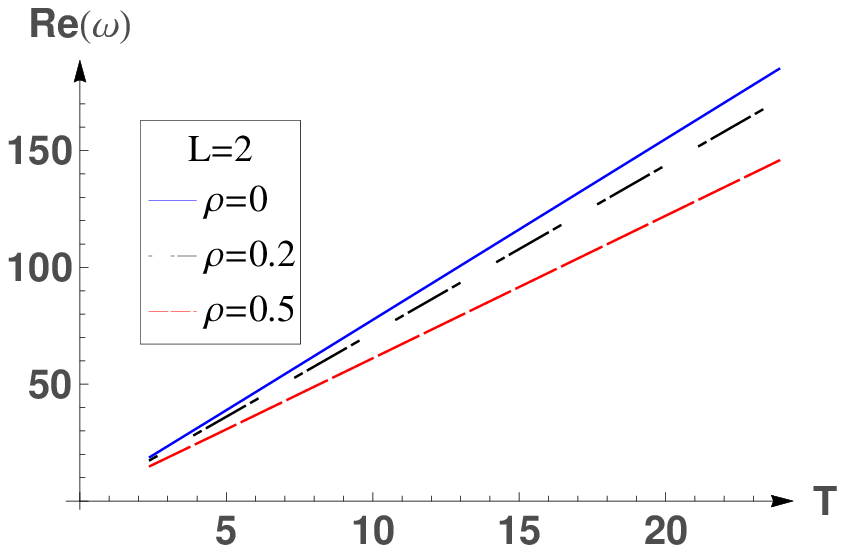}\protect\\
\includegraphics[width=5cm]{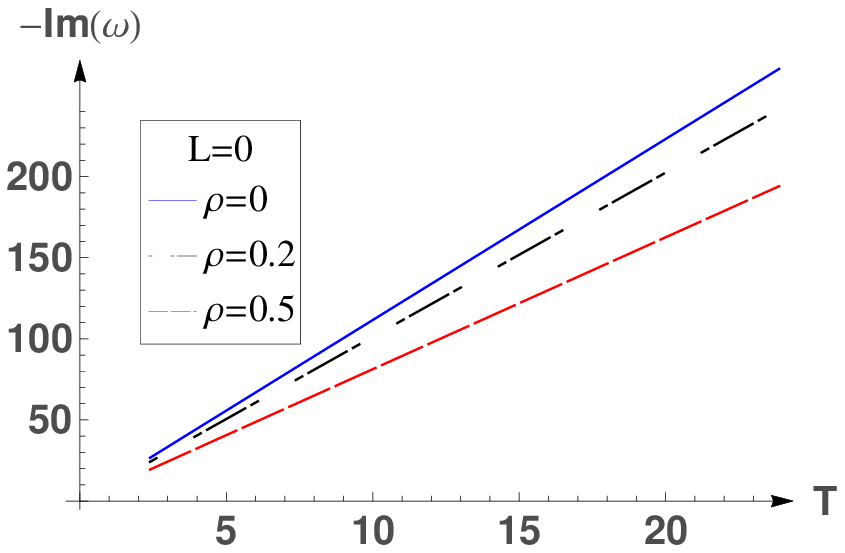}\includegraphics[width=5cm]{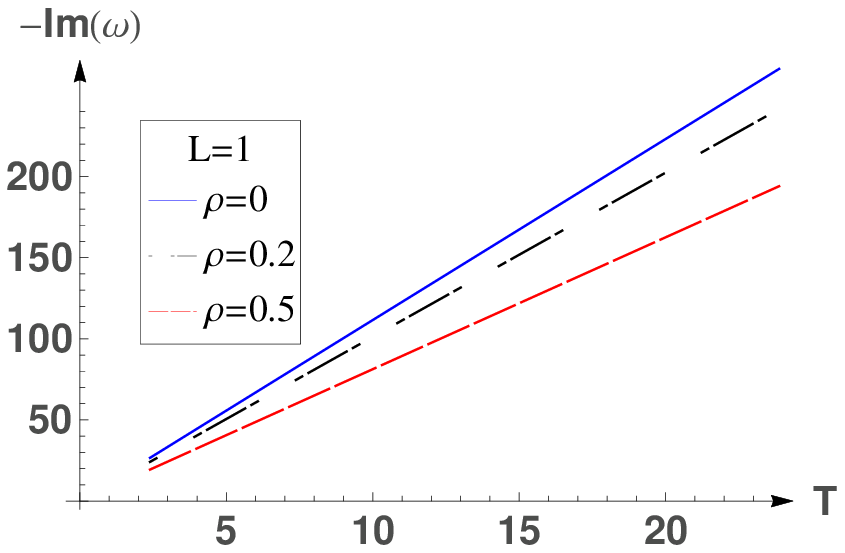}\includegraphics[width=5cm]{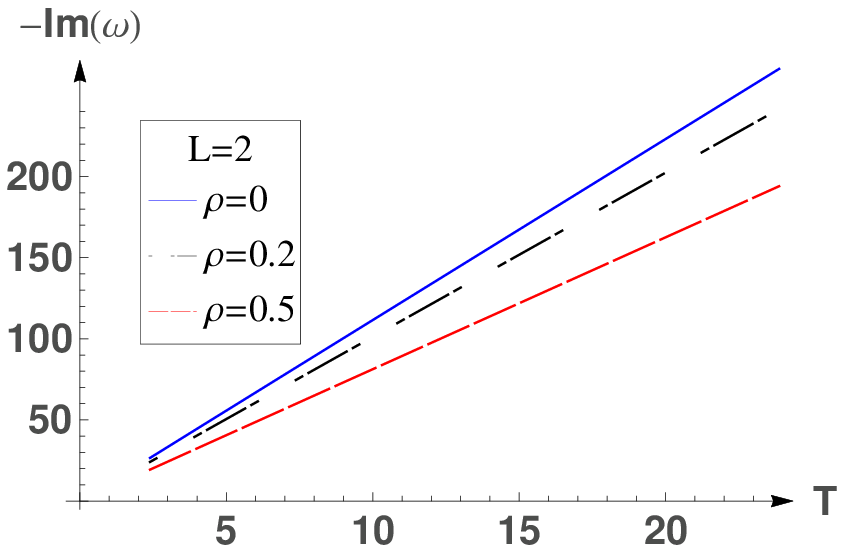}
\caption{Scalar quasinormal modes  vs. temperature of
Ho\v{r}ava-Lifshitz spherical black hole spacetime, where
$T=\frac{1+3r_0}{4\pi}$} \label{figB}
\end{figure*}

We show in Figs.(\ref{figA}, \ref{figB}) the results obtained by using Horowitz-Hubeny method.
By adopting the natural units, namely, $c=\hbar=G=k_B=1$, we express all physical quantities in terms of ``MeV". Therefore the units of temperature and frequency are both ``MeV".
In the Fig.\ref{figA} are shown the quasinormal modes for the planar black hole with $T=\frac{3r_0}{4\pi}$, while in Fig.(\ref{figB}) are presented the quasinormal modes for the spherical black hole where $T=\frac{1+3r_0}{4\pi}$.
It is found, from the above figures, that the quasinormal modes are mostly linear with respect to the temperature.
However, the slope of the line is affected by the correction (in terms of $\rho$) of Ho\v{r}ava-Lifshitz theory.
In particular, the correction increases the imaginary part of the frequency, but suppresses the real part.
Compared with the results obtained in general relativity, the quasinormal mode in Ho\v{r}ava-Lifshitz black holes have bigger period of quasinormal oscillation while its amplitude decays more slowly.

We summarize the quasinormal modes for the massless scalar field for some values of $L$ into Table \ref{TableI}.

\begin{table}[ht]
\caption{\label{TableI} The relation between $L$ and frequencies of Ho\v{r}ava-Lifshitz planar and spherical black hole spacetime with $\rho=0.1$ and $r_0=10$.}
\begin{tabular}{c c c}
         \hline
~~~$L$~~~&~~~$\omega$ (\text{Planar Black Hole})~~~&~~~$\omega$
(\text{Spherical Black Hole})~~~
        \\
        \hline
$0$&$17.8504-25.4472 i$&$17.9597-25.4498 i$
          \\
$1$&$17.9356-25.4244 i$&$18.0446-25.4270 i$
          \\
$2$&$18.1873-25.3571 i$&$18.2122-25.3818 i$
          \\
$3$&$18.5956-25.2492 i$&$18.4594-25.3158 i$
          \\
$4$&$18.9624-24.8185 i$&$18.7814-25.2307 i$
          \\
        \hline
\end{tabular}
\end{table}
We note that the smallest frequency occurs at $L=0$, which decays faster than the higher modes. A similar behaviour was pointed out by Horowitz and Hubeny in \cite{HH}. The imaginary part of the quasinormal modes are smaller in the planar black hole than in the spherical black hole. The real part of the quasinormal modes increases faster for the planar black hole than the spherical black hole after $L=2$.

 \section{Temporal Evolution}
\renewcommand{\theequation}{5.\arabic{equation}} \setcounter{equation}{0}

In this section, we employ the finite difference method \cite{FD} to study the temporal evolution of the quasinormal modes in the Ho\v{r}ava-Lifshitz black holes.
In this approach, one directly observes how small perturbations evolve in time.
To achieve this, we apply the finite difference method to Eq.(\ref{Fequation1}) and conveniently rescale the time, so that Eq.(\ref{Fequation1}) can now be rewritten as
 \bqn
 \lb{FequationT1}
\frac{\partial^2\Phi}{\partial {r_*^2}}-\frac{\partial^2\Phi}{\partial t^2}=V(r)\Phi.
 \eqn
By taking $t=t_0+i\Delta t$ and $r_*=r_{*0}+j\Delta r_*$ in Eq.(\ref{FequationT1}), the resulting finite difference equation reads
 \bqn
 \lb{FequationT2}
\Phi^{i+1}_j&=&-\Phi^{i-1}_j+\frac{\Delta t^2}{\Delta r_*^2}\left(\Phi^i_{j-1}+\Phi^i_{j+1}\right)\nb\\
&&+\left(2-2\frac{\Delta t^2}{\Delta r_*^2}-\Delta t^2V_j\right)\Phi^i_j.
 \eqn
The initial conditions are chosen to be
 \bqn
 \lb{FequationT3}
&&\Phi(r_*,t_0)=C_A\exp(-C_a(r_*-C_b)^2),\nb\\
&&\frac{\partial}{\partial t}\Phi(r_*,t)\Big|_{t=t_0}=0,
 \eqn
and the Dirichlet conditions at the Anti-de Sitter boundary is $\Phi(r_*,t)|_{r_*=0}=0$. In order to satisfy the Von Neumann stability
  \bqn
 \lb{FequationT4}
\frac{\Delta t^2}{\Delta r_*^2}+\frac{\Delta t^2}{4}V_{max}<1,
 \eqn
we choose $\Delta r_*=2\Delta t$, and $V_{max}\Delta t^2<3$, where $V_{max}$ is the largest value of $V_j$ in the numerical grids.

\begin{figure*}
\includegraphics[width=5cm]{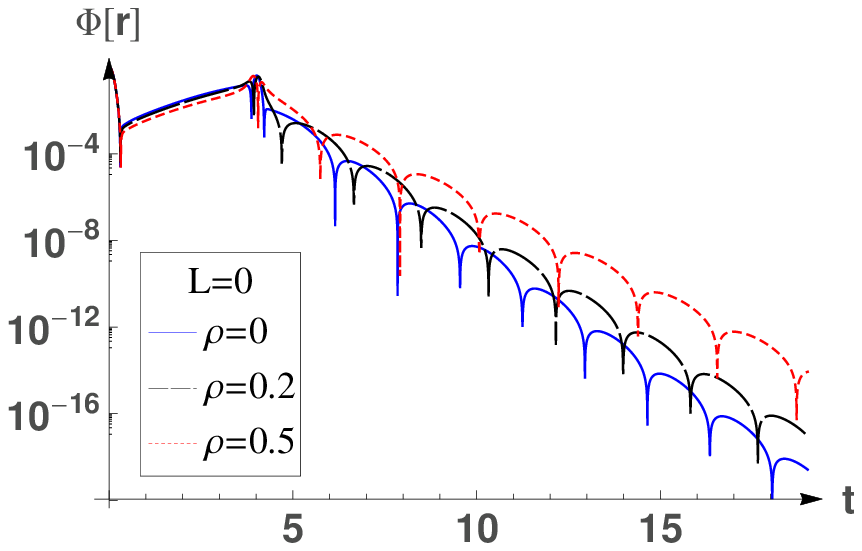}\includegraphics[width=5cm]{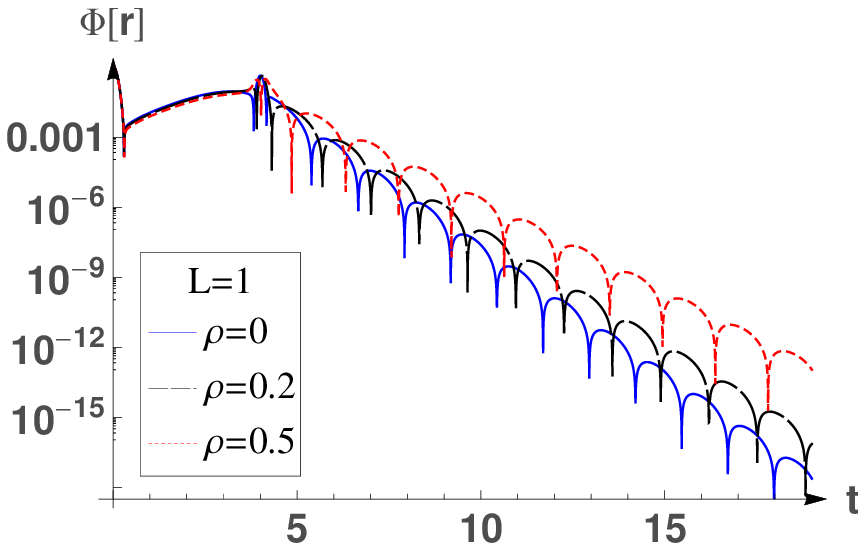}\includegraphics[width=5cm]{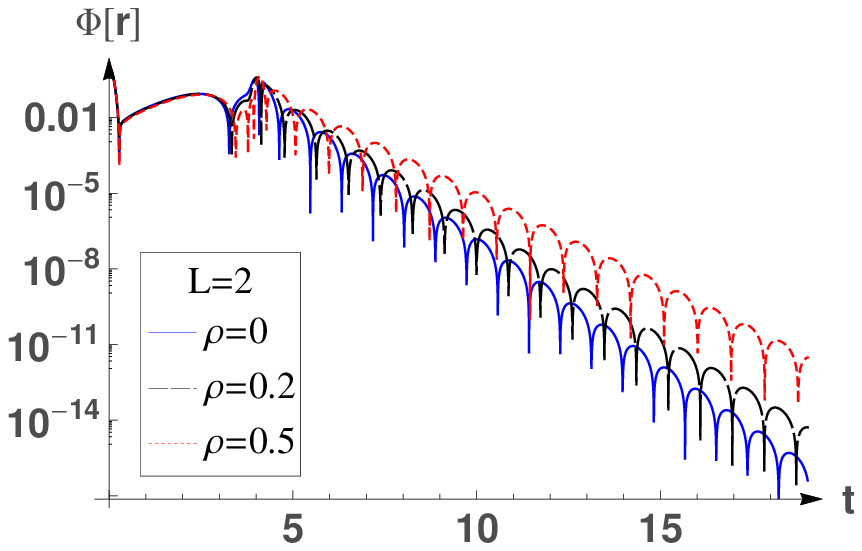}
\caption{The time evolutions of scalar perturbations in Ho\v{r}ava-Lifshitz planar black hole spacetime for $r_0=1$.}
\label{figC}
\end{figure*}

\begin{figure*}
\includegraphics[width=5cm]{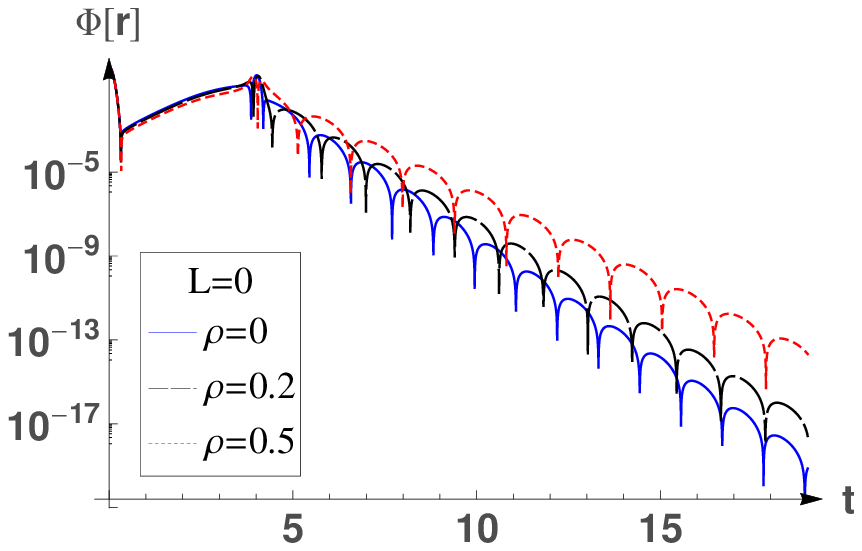}\includegraphics[width=5cm]{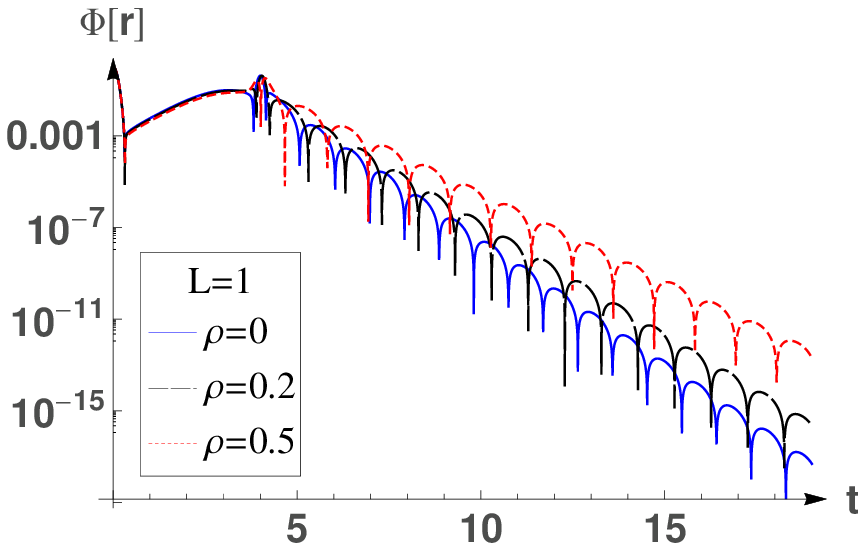}\includegraphics[width=5cm]{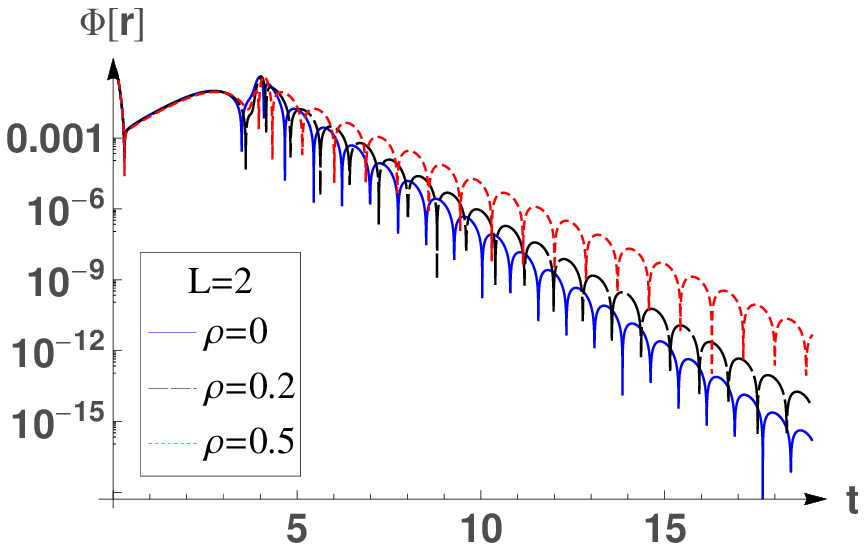}
\caption{The time evolutions of scalar perturbations in Ho\v{r}ava-Lifshitz spherical black hole spacetime for $r_0=1$.}
\label{figD}
\end{figure*}

The numerical results are presented in Figs.(\ref{figC}, \ref{figD}).
From the above plots, one may draw the same conclusion that the effect of the Ho\v{r}ava-Lifshitz spacetime is to increase the period of the quasinormal oscillation and its magnitude decays more slowly.

\section{Instability at large $\rho$}
\renewcommand{\theequation}{6.\arabic{equation}} \setcounter{equation}{0}

\begin{figure*}
\includegraphics[width=8cm]{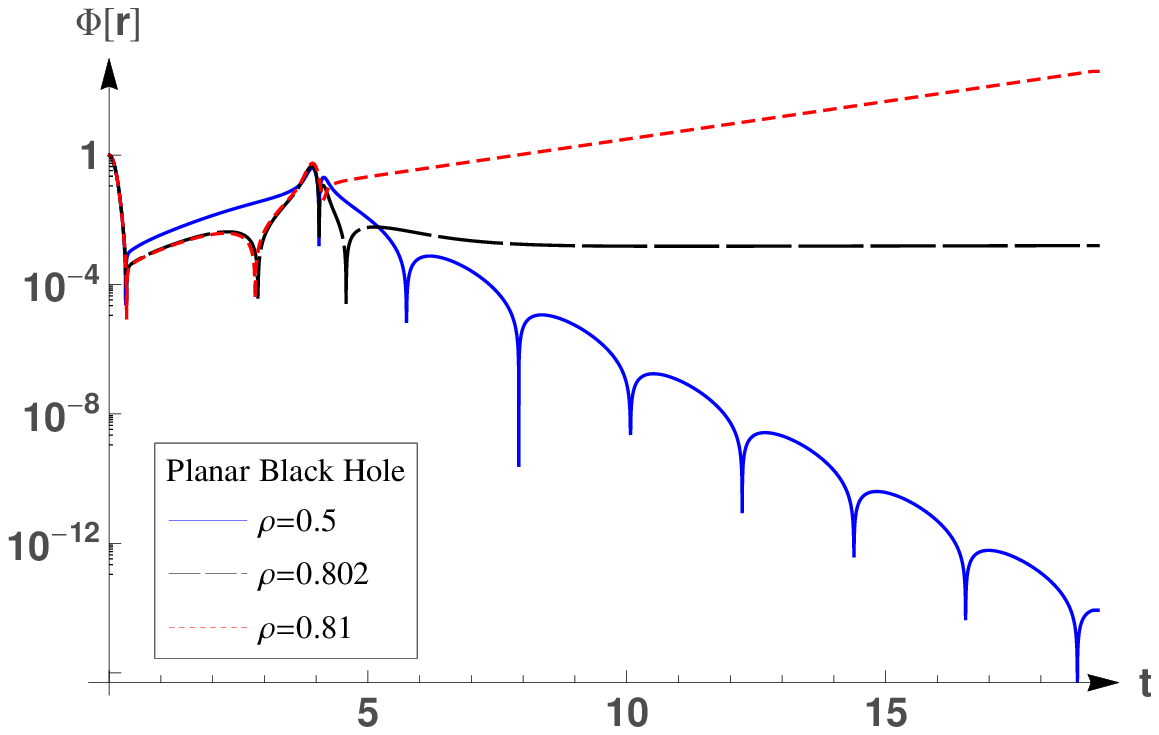}\includegraphics[width=8cm]{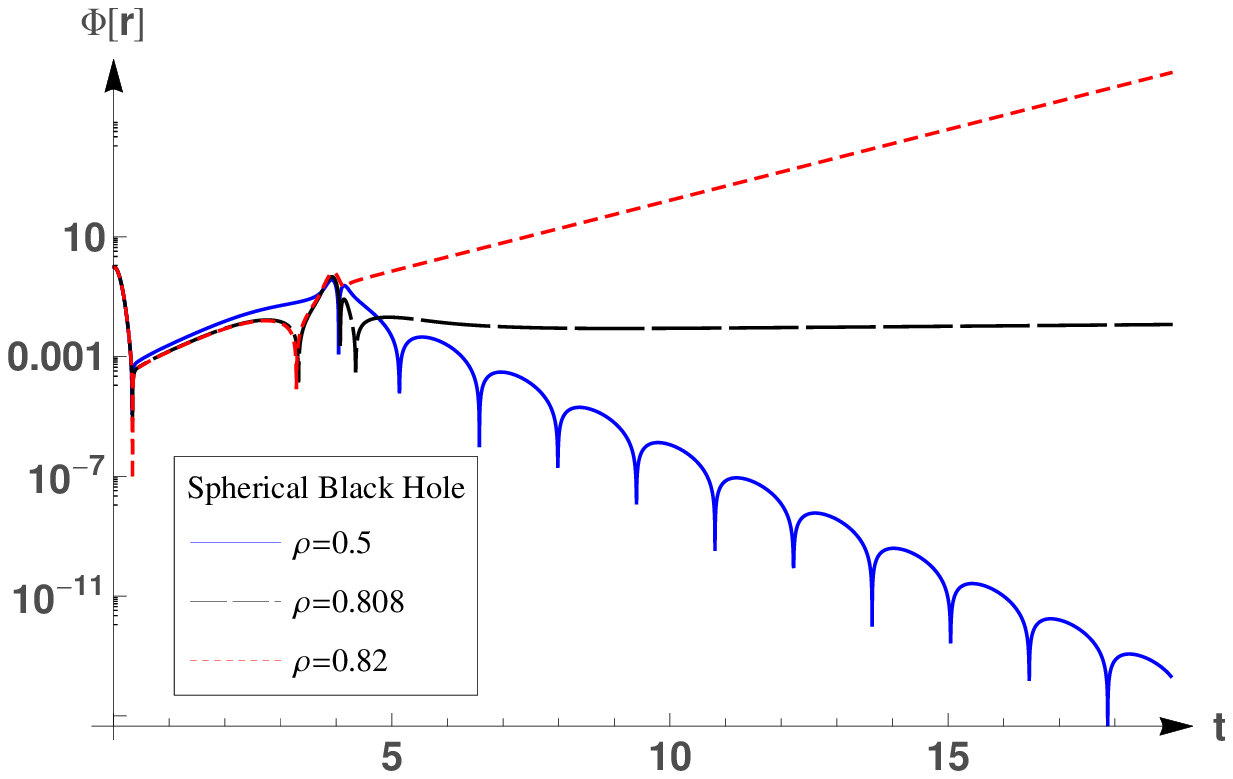}
\caption{Instability in planar and spherical black hole spacetime, for given $r_0=1$ and $L=0$.} \label{figE}
\end{figure*}

At last, we study numerically the stability of the scalar field perturbation in Ho\v{r}ava-Lifshitz spacetime in terms of $\rho$.
The results shown in Fig.(\ref{figE}) indicate that when $\rho$ becomes large enough, small perturbation may lead to instability of the system.
$\rho_c$, the critical value of $\rho$, is found numerically for a black hole with $r_0=1$ and $L=0$.
For planar black hole $\rho_c=0.802$ and for spherical black hole $\rho_c=0.808$.
Thus, in order to preserves the stability of the black holes against massless scalar perturbations, $\rho$ must be smaller than those critical values.

\begin{figure*}
\includegraphics[width=8cm]{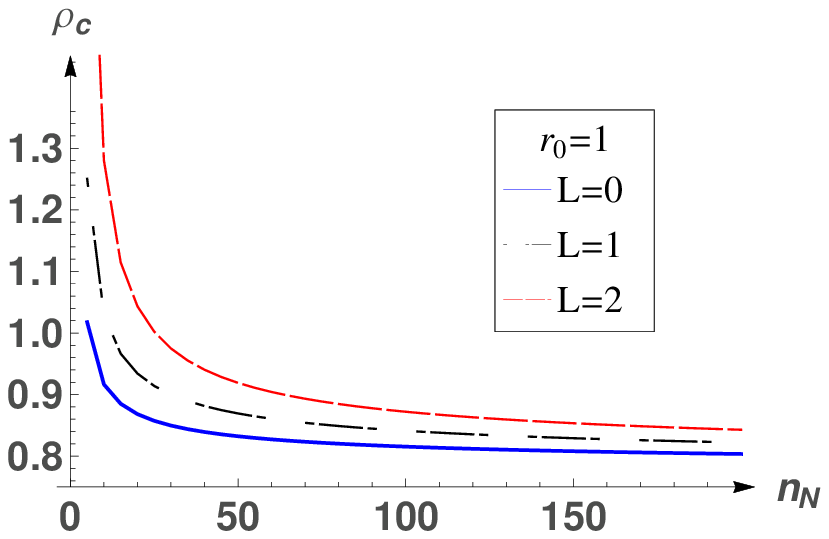}\includegraphics[width=8cm]{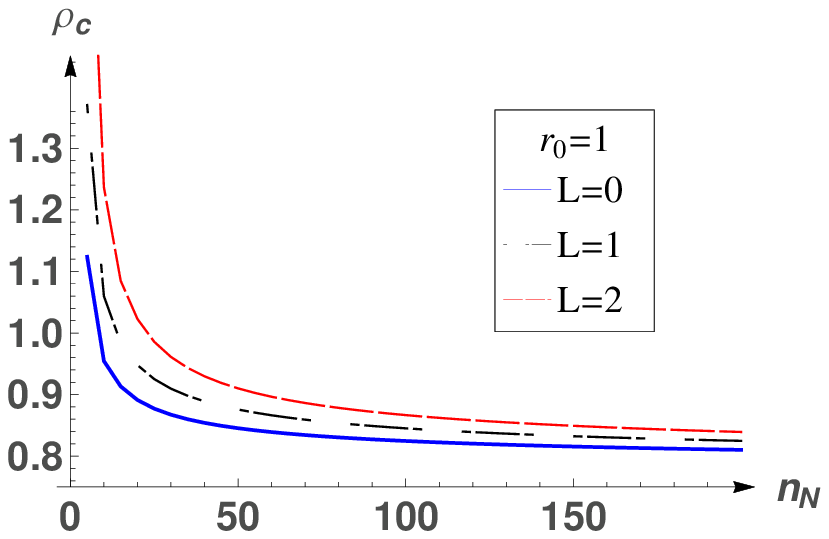}
\caption{The relation between the number of terms used in the expansion $n_N$ and the value of critical $\rho_c$ in planar (left hand side) and spherical (right hand side)
black hole spacetime, for given $r_0=1$.
The calculations are carried out by assuming $\omega=0$ for critical case.} \label{figF}
\end{figure*}

Now we proceed to study the frequencies of the quasi normal modes for the unstable region $\rho \ge \rho_c$.
For such region, one encounters some difficulties in terms of computational time when utilizing the Horowitz-Hubeny method to evaluate the quasinormal mode frequencies.
This is because in numerical calculations, one has to choose an integer $n_N$ as the number of terms used in the expansion of Eq.(\ref{FequationH6}).
This integer must be large enough so that the calculated frequency does not depend on the choice of $n_N$.
However, when $\rho$ increases and approaches its critical value $\rho_c$ from below, the minimal value of $n_N$ required for obtaining a convergent result tends to increase till the point that sometimes the numerical calculation becomes infeasible.
In Fig.\ref{figF}, we illustrate this fact by showing the relation between $n_N$ used in the calculation and the resultant $\rho_c$.
Since for the critical mode, one has $\omega=0$, we utilize this information to calculate the corresponding $\rho_c$.
Numerically, the calculation becomes less time consuming when one makes use of this condition,
though it is also possible to study the quasi normal mode without assuming $\omega=0$.
Since the calculation becomes very slow with large $n_N$, we will adopt the smallest possible value of $n_N$ while still achieving a reasonably good convergence.
From Fig.\ref{figF}, we find when $n_N \ge 50$, the results for $\rho_c$ become reasonable close to their convergent values.
Therefore we adopt $n_N=50$ in the following to evaluate the frequencies for critical as well as unstable modes.

\begin{figure*}
\includegraphics[width=8cm]{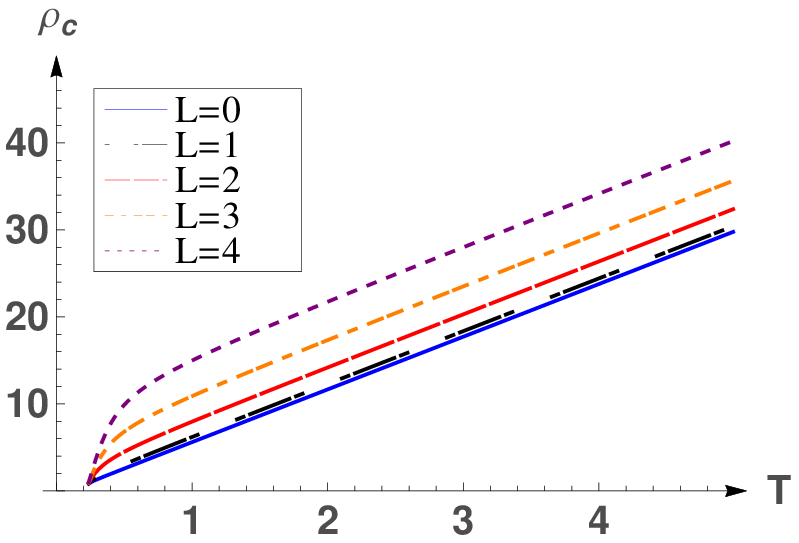}\includegraphics[width=8cm]{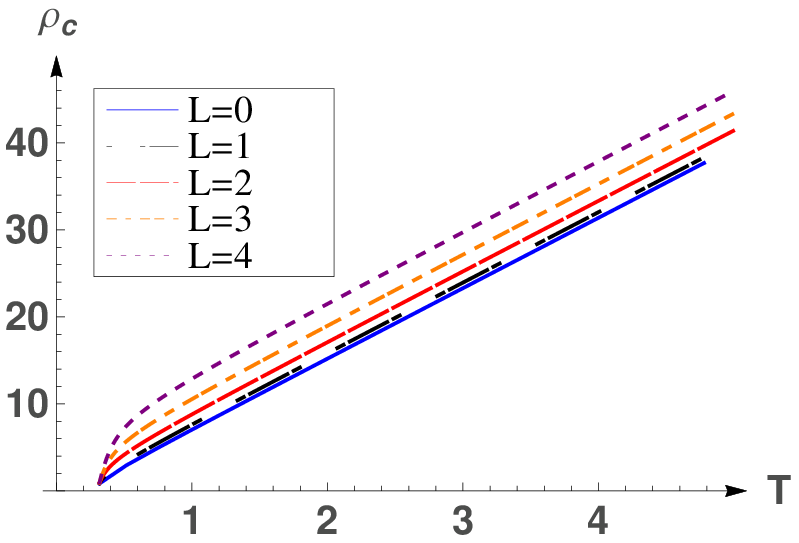}
\caption{The relation between the critical value $\rho_c$ and the temperature $T$ for planar (left hand side, $T=\frac{3r_0}{4\pi}$) and
spherical (right hand side, $T=\frac{1+3r_0}{4\pi}$) black hole spacetime.
For critical mode, the frequency $\omega$ vanishes.
The value $n_N=50$ is used in the calculations.} \label{figG}
\end{figure*}

In Fig.\ref{figG}, we show the value of $\rho_c$ as a function of the temperature $T$.
It is found that $T$ and $\rho_c$ becomes mostly linear at high temperature (or large $r_0$) region.
The frequencies of quasi normal modes in the unstable region are shown in Table \ref{TableII}.
It is confirmed that the real parts of the frequencies is numerically consistentely with zero in this case, and imaginary parts are positive.
Therefore the amplitudes of the oscillations are divergent when time goes to infinity, in accordance with the physical interpretation.

\begin{table}[ht]
\caption{\label{TableII} The relation between $\rho$ and frequencies
of the unstable quasi normal modes in Ho\v{r}ava-Lifshitz planar and spherical black hole spacetime with $L=0$ and $r_0=1$.
The number of terms in the expansion is $n_N=50$}
\begin{tabular}{c c c}
         \hline
~~~$\rho$~~~&~~~$\omega$ (\text{Planar Black Hole})~~~&~~~$\omega$
(\text{Spherical Black Hole})~~~
        \\
        \hline
$0.85$&$0.570335 i$&$0.241044 i$
          \\
$0.9$&$2.20793 i$&$2.15592 i$
          \\
$0.95$&$4.08224 i$&$4.07901 i$
          \\
        \hline
\end{tabular}
\end{table}

\section{Conclusion}
\renewcommand{\theequation}{7.\arabic{equation}} \setcounter{equation}{0}

In this work, we obtain the planar and spherical black hole solutions of Ho\v{r}ava-Lifshitz Gravity with local $U(1)$ symmetry, and investigate the massless scalar quasinormal modes in these spacetimes.
We find that the effect of Ho\v{r}ava-Lifshitz correction is to increases the period of quasinormal oscillation and slowing down the exponential decay.
Moreover, the scalar field's evolution will make the black hole unstable when $\rho$ becomes larger than a critical value.
The frequencies for critical and unstable modes are also evaluated.

As mentioned in \cite{UH}, even in the IR limit, some theory may
still allow the existence of massless particles with velocity
exceeding that of the light. For example, due to the presence of
spin-0 and spin-1 gravitons in Einstein-aether theory\cite{EA}, it
is required that these particles should propagate with a velocity no
less than that of the light, in order to cancel the Cherenkov
effects\cite{EMS}.

On the other hand, in UV case, Ho\v{r}ava-Lifshitz gravity allows
the existence of particles with infinite velocity. Our recent
studies on universal horizon show that even particles with
arbitrarily large velocities cannot escape from the inside of the
universal horizon\cite{UH,UHmore}. This implies that the universal
horizon, which situates inside of the event horizon, may play the
role of a real horizon of the black hole in a theory without Lorentz
symmetry. In this paper, we focus on low energy phenomena by
ignoring the higher energy terms in the action of the scalar field.
In this context, the boundary condition at event horizon shall be
valid, in other words, the event horizon plays the role of black
hole horizon.

We plan to investigate the quasinormal modes of Ho\v{r}ava-Lifshitz
black hole in UV limit. In this case, the black hole should be
defined by the universal horizon. In the UV limit, the
Horowitz-Hubeny method shall be modified accordingly because the
condition of a pure in-going mode at killing event horizon ceases to
be valid; while the finite difference method in Anti-de Sitter black
hole spacetime still works. In fact, in the finite difference
method, we find that the calculated perturbation won't reach the
killing event horizon, which situates outside of the universal
horizon. Nonetheless, besides the finite difference method, it is a
challenging and intriguing  task to further develop other methods in
order to study the properties of quasinormal modes of
Ho\v{r}ava-Lifshitz black hole in the UV limit.

Nevertheless, it is interesting to investigate the quasinormal modes of the matter field by taking into account those higher energy terms in Ho\v{r}ava-Lifshitz theory.
It is also compelling to investigate the gravitational perturbation, which in turn helps to understand the stability of the black holes and the properties of the gravitational wave in Ho\v{r}ava-Lifshitz theory.
We will devote ourselves to these topics in the future.

\section*{\bf Acknowledgements}

We would like to thank Prof. Anzhong Wang and Prof. Elcio Abdalla
for valuable discussions and insightful comments. This work is
supported in part by Brazilian funding agencies FAPESP, FAPEMIG,
CNPq and CAPES, and Chinese funding agency NNSFC under contract
No.11573022 and 11375279.

\section{Appendix: Projectable Planar Black Hole Solutions}
\renewcommand{\theequation}{A.\arabic{equation}} \setcounter{equation}{0}
\label{appendix}

In the Appendix, we derive a planar black hole solution with projectability condition and $U(1)$ symmetry. The projectable total action can be written as \cite{HLALW,HLHW},
 \bqn
\lb{A1}
S &=& \zeta^2\int dt d^{3}x N \sqrt{g} \Big({\cal{L}}_{K} -
{\cal{L}}_{{V}} +  {\cal{L}}_{{\varphi}} +  {\cal{L}}_{{A}} +  {\cal{L}}_{{\lambda}} \nb\\
& & ~~~~~~~~~~~~~~~~~~~~~~ \left. + {\zeta^{-2}} {\cal{L}}_{M} \right),
 \eqn
where ${\cal{L}}_{K}, \; {\cal{L}}_{\varphi}$ and ${\cal{L}}_{A}$ are given by Eq.(\ref{L0}) with
$a_i = 0$, and the potential ${\cal{L}}_{{V}}$ reads
 \bqn
 \lb{A2}
{\cal{L}}_{{V}} &=& 2\Lambda  + g_{1} R + \frac{1}{\zeta^{2}}
\left(g_{2}R^{2} +  g_{3}  R_{ij}R^{ij}\right)\nb\\
& & + \frac{1}{\zeta^{4}} \left(g_{4}R^{3} +  g_{5}  R\;
R_{ij}R^{ij}
+   g_{6}  R^{i}_{j} R^{j}_{k} R^{k}_{i} \right)\nb\\
& & + \frac{1}{\zeta^{4}} \Big[g_{7}(\nabla R)^{2} +  g_{8}
\left(\nabla_{i}R_{jk}\right) \left(\nabla^{i}R^{jk}\right)\Big],
~~~~
 \eqn
where the coupling  constants $ g_{s}\, (s=1, 2,\dots 8)$  are all dimensionless, and we set $g_1=-1$ in the following calculations (In fact, we find the form of ${\cal L}_V$ only influences the solution of $A(r)$ in spherically symmetric spacetime).

The planar black hole metric is given by
  \bqn
  \lb{A3}
  ds^2=-N^2dt^2+\frac{\left[dr+h(r)dt\right]^2}{f(r)}+r^2(dx^2+dy^2),
  \eqn
since the projectability condition requires $N=N(t)$, we set $N=1$ without loss the generality. Subsequently, the field equations can be derived from the action (\ref{A1}) \cite{HLALW}. We pick the gauge $\varphi=0$, and obtain the following field equations
 \bqn
 \lb{A4}
rf'+f+\Lambda_g r^2=0,
 \eqn
 \bqn
 \lb{A5}
(\lambda-1)\left[h''+\left(\frac{2}{r}-\frac{f'}{2f}\right)h'\right]+\frac{f'h}{rf}&&\nb\\
+(\lambda-1)\left(\frac{f'^2}{2f^2}+\frac{f''}{2f}-\frac{2}{r^2}\right)h&=&0,
 \eqn
 \bqn
 \lb{A6}
&&A'+\left(\frac{1}{r}+\frac{r\Lambda_g}{f}\right)\frac{A'}{2}-\frac{\Lambda r^2+f}{2rf}+\frac{hh'}{f}\nb\\
&&+\left(\frac{1}{2rf}-\frac{f'}{f^2}\right)h^2+(\lambda-1)\Bigg[\frac{rh'^2}{4f}-\frac{rhh''}{2f}\nb\\
&&+\left(\frac{2}{rf}-\frac{f'}{2f^2}-\frac{3rf'^2}{16f^3}+\frac{rf''}{4f^2}\right)h^2\Bigg]+Q(r)=0,\nb\\
 \eqn
where $Q(r)$ is higher energy term, which is given by
 \bqn
 \lb{A7}
Q(r)&=&\frac{(8g_2+3g_3)(r^2\Lambda_g+2f)r^2\Lambda_g-g_3f^2}{8\xi^2r^3f}\nb\\
&&+\frac{1}{8\xi^4r^5f}\Big[(32g_4+12g_5+5g_6-g_8)\Lambda_g^3r^6\nb\\
&&+(48g_4+14g_5+3g_6+2g_8)\Lambda_g^2r^4f\nb\\
&&-(40g_4+45g_6-35g_8)\Lambda_gr^2f^2\nb\\
&&-(66g_5+75g_6-60g_8)f^3\Big]
 \eqn
From Eq.(\ref{A4}), we get
 \bqn
 \lb{A8}
f(r)=\frac{f_0}{r}-\frac{\Lambda_g}{3}r^2,
 \eqn
where $f_0$ is a constant. Then, one can use Eq.(\ref{A5}) and Eq.(\ref{A6}) to calculate $h(r)$ and $A(r)$.

\subsection{$\lambda=1$ Case}

When $\lambda=1$, from Eq.(\ref{A5}), we have $h(r)=0$. By substituting $h=0$ into Eq.(\ref{A6}), one obtains
 \bqn
 \lb{A9}
A(r)&=&\sqrt{f(r)} A_0+\frac{\Lambda }{\Lambda _g}-\frac{3 \left(22 g_5+25 g_6-20 g_8\right) f(r)^2}{44 \xi ^4 r^4}\nb\\
&&-\Lambda_g \left[\frac{242 g_5+270 g_6-205 g_8}{220 \xi^4 r^2}f+\frac{8 g_2+3 g_3}{4 \xi ^2}\right]\nb\\
&&-\frac{g_3 f}{20 \xi ^2 r^2}+\frac{\left(-32 g_4-12 g_5-5 g_6+g_8\right) \Lambda _g^2}{4 \xi ^4}\nb\\
&&-A_S(r) \Big[\frac{2 \left(110 g_4+44 g_5+20 g_6-5 g_8\right) \Lambda_g^3}{55 \xi ^4}\nb\\
&&+\frac{2 \left(5 g_2+2 g_3\right) \Lambda _g^2}{5 \xi^2}-\Lambda _g+\Lambda \Big]
 \eqn
where $A_0$ is a constant, and
 \bqn
 \lb{A10}
&&A_S(r)=\frac{i(-1)^{5/12}f_0^{1/2}f(r)^{1/2}}{3^{1/12}(-f_0)^{5/6}(-\Lambda_g)^{7/6}}\nb\\
&&\times F\left[\arcsin\left(3^{-\frac14}\sqrt{-(-1)^{\frac56}-\frac{i3^{\frac13}(-f_0)^{\frac13}}{r(-\Lambda_g)^{\frac13}}}\right)\Bigg|(-1)^{\frac13}\right],\nb\\
 \eqn
and $F(\phi|m)=\int^\phi_0[1-m\sin^2(\theta)]^{-1/2}d\theta$ is the elliptic integral.

In particular, when $\Lambda=\Lambda_g$ and $g_i=0$, we find
 \bqn
 \lb{A11}
A(r)=1+A_0\sqrt{f(r)}
 \eqn
where $A_0$ is a constant.

\subsection{$\lambda\not=1$ and $f_0=0$ Case}

In this case, we find
 \bqn
 \lb{A12}
h&=&h_1r^{\sqrt{\frac{\lambda-3}{\lambda-1}}}+h_2r^{-\sqrt{\frac{\lambda-3}{\lambda-1}}},\nb\\
A&=&A_0r-\frac{1}{2}+\frac{3\Lambda}{2\Lambda_g}-\frac{2\Lambda_g^2}{3\xi^4}(9g_4+3g_5+g_6)\nb\\
&&-\frac{\Lambda_g}{3\xi^2}(3g_2+g_3)+\frac{3h_1^2}{2\Lambda_gr^{2\left(1+\sqrt{\frac{\lambda-3}{\lambda-1}}\right)}}\nb\\
&&\times\Bigg[\frac{\sqrt{(\lambda-1)^3(\lambda-3)}+2\sqrt{(\lambda-1)(\lambda-3)}}{3\sqrt{\lambda-1}+2\sqrt{\lambda-3}}\nb\\
&&-\frac{(\lambda-1)(\lambda-3)}{3\sqrt{\lambda-1}+2\sqrt{\lambda-3}}\Bigg]+\frac{3h_2^2}{2\Lambda_gr^{2\left(-1+\sqrt{\frac{\lambda-3}{\lambda-1}}\right)}}\nb\\
&&\times\Bigg[\frac{\sqrt{(\lambda-1)^3(\lambda-3)}+2\sqrt{(\lambda-1)(\lambda-3)}}{-3\sqrt{\lambda-1}+2\sqrt{\lambda-3}}\nb\\
&&+\frac{(\lambda-1)(\lambda-3)}{-3\sqrt{\lambda-1}+2\sqrt{\lambda-3}}\Bigg]
 \eqn
where $h_1$ and $h_2$ are constants.

\subsection{$\lambda\not=1$ and $\Lambda_g=0$ Case}

In this case, we find
 \bqn
 \lb{A13}
h&=&h_1r^{-\frac34-\Lambda_C}+h_2r^{-\frac34+\Lambda_C},\nb\\
A&=&\frac{A_0}{\sqrt{r}}+1+\frac{\Lambda r^3}{7f_0}+\frac{h_2^2}{8f_0r^{\frac{1}{2}+2\Lambda_C}}\left(\frac{1}{\Lambda_C}-4\right)\nb\\
&&-\frac{h_1^2}{8f_0r^{\frac{1}{2}+2\Lambda_C}}\left(\frac{1}{\Lambda_C}+4\right)+\frac{h_1h_2\ln(r)}{2f_0\sqrt{r}}\nb\\
&&-\frac{3f_0^2}{44\xi^4r^6}(22g_5+25g_6-20g_8)-\frac{g_3f_0}{20\xi^2r^3}\nb\\
&&+\frac{\lambda-1}{128f_0r^{\frac{1}{2}+2\Lambda_C}}\Big[4h_1h_2r^{2\Lambda_C}(49-\Lambda_C^2)\ln(r)\nb\\
&&+\left(48\Lambda_C-56-\frac{49}{\Lambda_C}\right)h_1^2\nb\\
&&-\left(48\Lambda_C+56-\frac{49}{\Lambda_C}\right)h_2^2r^{4\Lambda_C}\Big],
 \eqn
where $\Lambda_C=\frac{1}{4}\sqrt{\frac{49\lambda-33}{\lambda-1}}$.

\subsection{General Case}
In a general case, $A(r)$ should be obtained from Eq.(\ref{A6}), namely,
 \bqn
 \lb{A14}
A(r)=A_0\sqrt{f}-\sqrt{f}\int \frac{Q_A(r)}{\sqrt{f}} dr,
 \eqn
where
 \bqn
 \lb{A15}
Q_A(r)&=&Q(r)-\frac{\Lambda r^2+f}{2rf}+\frac{hh'}{f}\nb\\
&&+\left(\frac{1}{2rf}-\frac{f'}{f^2}\right)h^2+(\lambda-1)\Bigg[\frac{rh'^2}{4f}-\frac{rhh''}{2f}\nb\\
&&+\left(\frac{2}{rf}-\frac{f'}{2f^2}-\frac{3rf'^2}{16f^3}+\frac{rf''}{4f^2}\right)h^2\Bigg].
 \eqn
To calculate $h(r)$, one makes use of Eq.(\ref{A5}). We introduce the following transformation
 \bqn
 \lb{A16}
 x&=&\frac{\Lambda_g^{1/3}r}{3^{1/3}f_0^{1/3}},\nb\\
 h&=&x^{\frac{2c-5}{4}(1-x^3)^{\frac{3+2b-2c}{12}}}H(x),
 \eqn
and Eq.(\ref{A5}) becomes
 \bqn
 \lb{A17}
x(1-x^3)H''(x)+(c-bx^3)H'(x)+ax^2H(x)=0\nb\\
 \eqn
where the $a$, $b$ and $c$ may be given by
 \bqn
 \lb{A18}
a&=&\frac{3+21\lambda}{8-8\lambda}-\frac{3}{2}\Lambda_C,~~b=\frac{5}{2}+2\Lambda_C,\nb\\
c&=&1+2\Lambda_C,
 \eqn
or
 \bqn
 \lb{A19}
a&=&\frac{33-57\lambda}{8\lambda-8}-\frac{9}{2}\Lambda_C,~~b=\frac{11}{2}+2\Lambda_C,\nb\\
c&=&1+2\Lambda_C,
 \eqn
or
 \bqn
 \lb{A20}
a&=&\frac{3+21\lambda}{8\lambda-8}+\frac{3}{2}\Lambda_C,~~b=\frac{5}{2}-2\Lambda_C,\nb\\
c&=&1-2\Lambda_C,
 \eqn
or
 \bqn
 \lb{A21}
a&=&\frac{33-57\lambda}{8\lambda-8}+\frac{9}{2}\Lambda_C,~~b=\frac{11}{2}-2\Lambda_C,\nb\\
c&=&1-2\Lambda_C,
 \eqn
If we set $y=x^3$, Eq.(\ref{A17}) becomes a hypergeometric equation
 \bqn
 \lb{A22}
y(1-y)\frac{d^2H}{dy^2}+\frac{1}{3}\left[2+c-(2+b)y\right]\frac{dH}{dy}+\frac{a}{9}H=0,\nb\\
 \eqn
and the solution is
 \bqn
 \lb{A23}
H&=&\frac{h_1}{y^{\frac{1-c}{3}}}~~{}_2F_1\left(H_{A-},H_{A+};\frac{4-c}{3};y\right)\nb\\
&&+h_2~~{}_2F_1\left(H_{B-},H_{B+};\frac{2+c}{3};y\right),\nb\\
&=&\frac{h_1}{x^{1-c}}~~{}_2F_1\left(H_{A-},H_{A+};\frac{4-c}{3};x^3\right)\nb\\
&&+h_2~~{}_2F_1\left(H_{B-},H_{B+};\frac{2+c}{3};x^3\right),
 \eqn
where $H_{A\pm}=\frac{b+1}{6}-\frac{c}{3}\pm\frac{1}{6} \sqrt{4 a+b^2-2b+1}$, $H_{B\pm}=\frac{b-1}{6}\pm\frac{1}{6} \sqrt{4 a+b^2-2b+1}$, and ${}_2F_1(a,b;c;d)$ is the hypergeometric function.

\onecolumngrid

\end{document}